\documentclass[showpacs,pre,twocolumn,amsmath,amssymb,epsfig]{revtex4-1}

\usepackage{graphicx}
\usepackage{psfrag}
\usepackage{epsfig}
\usepackage{dcolumn}
\usepackage{bm}
\newcommand     {\beq}[1]         { \begin{equation} #1 \end{equation} }

\begin{document}

\title{Creep rupture as a non-homogeneous Poissonian process}

\author{Zsuzsa Danku}
\affiliation{Department of Theoretical Physics, University of Debrecen,
P.O. Box 5, H-4010 Debrecen, Hungary}

\author{Ferenc Kun}
\email{ferenc.kun@science.unideb.hu}
\affiliation{Department of Theoretical Physics, University of Debrecen,
P.O. Box 5, H-4010 Debrecen, Hungary}

\begin{abstract}
Creep rupture of heterogeneous materials occurring under constant sub-critical
external loads is responsible for the collapse of engineering constructions and
for natural catastrophes. Acoustic monitoring of crackling bursts provides
microscopic insight into the failure process.  Based on a fiber bundle model, we
show that the accelerating bursting activity when approaching failure can be
described by the Omori law. For long range load redistribution the time series
of bursts proved to be a non-homogeneous Poissonian process with power law
distributed burst sizes and waiting times. We demonstrate that limitations of
experiments such as finite detection threshold and time resolution have striking
effects on the characteristic exponents, which have to be taken into account
when comparing model calculations with experiments. Recording events solely
within the Omori time to failure the size distribution of bursts has a crossover
to a lower exponent which is promising for forecasting the imminent catastrophic
failure.
\end{abstract}

\maketitle

Materials subject to a constant external load below their fracture strength
typically exhibit a time dependent response and fail in a finite time. Such
creep rupture phenomena have an enormous technological importance and human
impact since they are responsible for the collapse of engineering constructions
and they lie at the core of natural catastrophes such as landslides, snow and
stone avalanches and earthquakes \cite{1,2,3,4,5,6,7,8,9,10,11}. The acoustic
waves generated by the nucleation and propagation of cracks allow for the
monitoring of the failure process on the meso- and micro scales. Crackling noise
is usually characterized by the integrated statistics accumulating all the
events of the time series up to failure \cite{2,3,4,5,6,7,8,9,10,11,12,13,14}.
Experiments revealed that the probability distribution of the energy of
crackling bursts and of the interoccurrence times have power law functional
form, which are considered to be the fingerprint of correlations in the
microscopic breaking dynamics \cite{1,2,3,4,5,6,7,8,9,10,11,12,13,14}. The value
of the exponents measured on different types of heterogenous materials show a
surprisingly large scatter between $1$ and $2$ which has not been captured by
theoretical studies \cite{1,2,3,4,5,6,7,8,9,10}. The approach to failure is
usually characterized on the macroscale by the strain rate which proved to have
a power law divergence as a function of time to failure \cite{1,5,6,14}.

\begin{figure}
\begin{center}
\epsfig{bbllx=40,bblly=0,bburx=640,bbury=480,
file=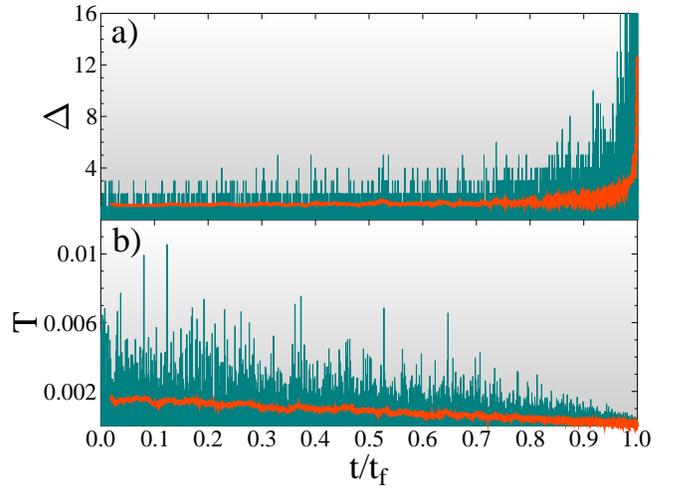, width=8.3cm}  
\caption{Time series of bursts in a bundle of $N=100000$ fibers at a load
$\sigma_0 / \sigma_c = 0.05$: the size of bursts $\Delta$ $(a)$ and waiting
times $T$ between consecutive events $(b)$ are presented as function of time $t$
of their occurrence normalized by the lifetime $t_f$ of the system. The red
lines represent the moving average of $\Delta$ and $T$. The increasing average
burst size and decreasing average waiting time indicate the acceleration of the
system towards failure.
}
\label{fig:timeseries}
\end{center}
\end{figure}

Here we take a different strategy and investigate the details of the crackling
time series in order to understand how the creeping system evolves towards
catastrophic failure. We consider a generic fiber bundle model
\cite{16,17,18,19,20} (FBM) of damage enhanced creep rupture which successfully
reproduces measured creep behaviour of heterogeneous materials both on the micro
and macro scales (Methods) \cite{21,22}. In the model under a constant
subcritical external load the fibers break due to two physical mechanisms:
immediate breaking occurs when the load of fibers exceeds the local failure
strength. Time dependence emerges such that intact fibers accumulate damage
which results in failure in a finite time. The separation of time scales of slow
damaging and of immediate breaking together with the load redistribution
following failure events lead to a highly complex time evolution where slowly
proceeding damage sequences trigger bursts of immediate breakings \cite{21,22}.
An example of the time series of bursts can be seen in Fig.\
\ref{fig:timeseries} where the increasing burst size $\Delta$ and the decreasing
waiting time $T$ between consecutive events clearly mark the acceleration of the
system towards failure. As a novel approach to creep we focus on the evolution
of the rate of bursts and show that the temporal occurrence of crackling events
and the power law statistics of waiting times can fully be described based on
non-homogeneous Poissonian processes without assuming correlations of bursts.
Our investigation unveils that limitations of measuring devices in experiments
have astonishing effects on the outcomes of crackling noise analysis which can
explain the strong scatter of measured critical exponents of crackling noise in
creep, and the discrepancy between experimental findings and theoretical
approaches. Studying how the time series evolves when approaching the
catastrophe we address the possibility of forecasting the imminent failure. The
importance of the results goes beyond fracture phenomena and catastrophic
failures, recently the human activity has been found to exhibit similar bursty
character where analogous problems of the evolution of time series and waiting
time statistics occur \cite{23,24,25}. 

\section{Results}
Bursts of immediate breakings triggered by damage sequences are analogous to
acoustic outbreaks in loaded specimens. However, damage breakings cannot be
recorded by experimental means they determine the waiting time $T$ between
consecutive bursts. The competition of the two failure modes has the consequence
that the system drives itself towards failure under a constant subcritical
external load $\sigma_0 < \sigma_c$. The global acceleration of the dynamics
that can be observed in Fig.\ \ref{fig:timeseries} is the consequence of the
increasing load on the intact part of the system due to subsequent load
redistributions, while the fluctuations of the burst size $\Delta$ and waiting
time $T$ emerge due to the quenched heterogeneity of fibers’ strength in
qualitative agreement with experiments \cite{1,2,3,4,5,6,7,8,9,10,11,12,13}.

\begin{figure}
\begin{center}
\epsfig{bbllx=30,bblly=470,bburx=760,bbury=790,
file=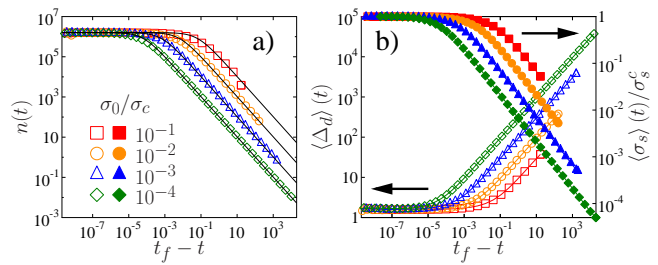, width=8.2cm}  
\caption{$(a)$ Event rate $n$ as a function of time to failure $t_f-t$ for
several different load values. The continuous lines represent fits with the
Omori law equation (\ref{eq:omori}). $(b)$ Average size of damage sequences
$\left<\Delta_d\right>$ and average load of single fibers $\sigma_s$ normalized
by its quasi-static critical value $\sigma_s^c$ as function of time to failure.}
\label{fig:omori_illesztes}
\end{center}
\end{figure}

In order to quantify how the accelerating dynamics appears on the microscale we
determined the rate of bursts $n\left(t\right)$ as a function of the distance
from the critical point $t_f-t$. Fig.\ \ref{fig:omori_illesztes}$(a)$ shows that
at the beginning of the creep process the event rate monotonically increases
having a power law functional form. Approaching catastrophic failure
$n\left(t\right)$ saturates and converges to a constant. The most remarkable
result is that the functional form of $n\left(t\right)$ can be described by the
modified Omori law \cite{26,27}
\beq{
n(t) = \displaystyle{\frac{A}{\left[1+(t_f-t)/c\right]^p}},
\label{eq:omori}
}
where $A$ is the saturation rate or productivity at catastrophic failure, $c$
denotes the characteristic time scale, and $p$ is the Omori exponent. Perfect
agreement can be observed in Fig.\ \ref{fig:omori_illesztes}$(a)$ between the
simulated data and the analytic form of equation (\ref{eq:omori}). In the case
of earthquakes, the Omori law describes the decay rate of aftershocks following
major earthquakes \cite{26,27}. For creep rupture we observe the inverse
process: considering the macroscopic failure as the main shock, the breaking
bursts are foreshocks whose increasing rate is described by the (inverse) Omori
law.

As a crucial point, our approach makes it posssible to clarify how the
characteristic time scale $c$ of the Omori law emerges: Fig.\
\ref{fig:omori_illesztes}$(b)$ illustrates that due to the increasing load on
intact fibers shorter and shorter damage sequences are sufficient to trigger
bursts. However, this acceleration is limited such that for $t_f-t<c$ the
average length of damage sequences $\left<\Delta_d\right>\left(t\right)$
saturates between $1$ and $2$. The origin of this high susceptibility is that
the load of intact fibers $\sigma_s\left(t\right)$ gradually increases to its
quasi-static critical value $\sigma_s^c$ (see Fig.\
\ref{fig:omori_illesztes}$(b)$) \cite{16,17,18,19,20,21,22}. Hence, the Omori
time scale $c$ is determined by the condition
$\sigma_s\left(t_f-t=c\right)\approx\sigma_s^c$, where $\sigma_s^c$ can be
obtained from the quasi-static constitutive equation of FBMs
\cite{16,17,18,19,20,21,22}. Note that the condition $\left<\Delta_d\right>=1$
marks the point of instability where the avalanche cannot stop anymore and it
becomes catastrophic. 

Fig.\ \ref{fig:omori_illesztes}$(a)$ also demonstrates that the saturation rate
$A$ does not depend on the external load $\sigma_0$, however, the characteristic
time scale $c$ linearly increases $c \sim \sigma_0$, indicating that at higher
load saturation sets on earlier. Our simulations revealed that the Omori
exponent is $p=1$, it does not depend on any details of the damage law
\cite{21,22} such as the $\gamma$ exponent and the disorder distributions until
the load redistribution is long ranged.

\begin{figure}
\begin{center}
\epsfig{bbllx=0,bblly=480,bburx=360,bbury=780, 
file=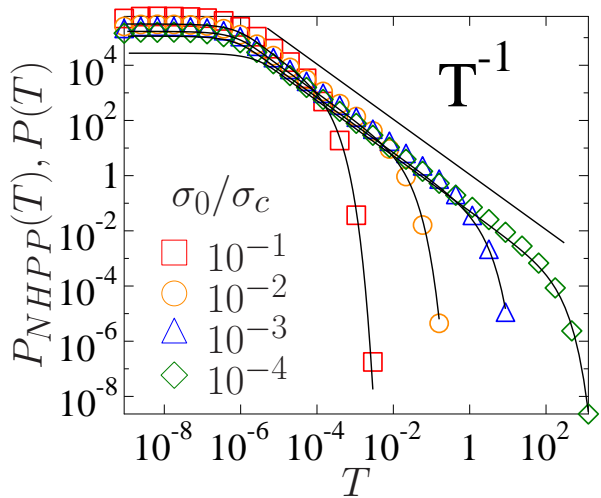, width=8.0cm}
\caption{Probability distribution of waiting times $P\left(T\right)$ between
consecutive bursts. The numerical results (symbols) are perfectly described by
the analytic prediction $P_{NHPP}\left(T\right)$ of equation
(\ref{eq:wait_nhpp}) (continuous lines).}
\label{fig:waiting_time}
\end{center}
\end{figure}

The event rate $n\left(t\right)$ is practically the inverse of the average
waiting time between events occurring at time $t$. More detailed
characterization is provided by the probability distribution of waiting times
$P\left(t\right)$ which is presented in Fig.\ \ref{fig:waiting_time}
corresponding to the system of Fig.\ \ref{fig:omori_illesztes}. Along the
distributions two characteristic time scales can be identified: for waiting
times below a threshold $T<T_l$, the distributions have constant values, while
in the limit of large waiting times $T>T_u$ a rapidly decreasing exponential
form is obtained.

For the intermediate regime $T_l<T<T_u$ the waiting time distributions exhibit a
power law behavior
\beq{
P(T) \sim T^{-z},
} 
where the exponent proved to be universal $z=1$. Increasing the external load
$\sigma_0$ the upper cutoff $T_u$ decreases, however, the lower characteristic
time $T_l$ is independent of $\sigma_0$. Since the temporal occurrence of events
is determined by the global increase of the breaking probability due to the
increasing load on intact fibers, the above results suggest that the time
evolution of crackling noise of heterogeneous materials undergoing creep rupture
can be described as a non-homogeneous Poissonian process (NHPP). For NHPP the
waiting time distribution of a series of $N_{\Delta}$ events with duration $t_f$
can be obtained analytically from the event rate as \cite{28,29}
\begin{eqnarray}
P_{NHPP}(T) &=&
\frac{1}{N_{\Delta}}\int_{0}^{t_f-T}n(s)n(s+T)e^{-\int_s^{s+T}n(u)du}ds 
\nonumber\\
&+& n(T)e^{-\int_0^Tn(s)ds}.
\label{eq:wait_nhpp}
\end{eqnarray}

To verify the consistency of the NHPP picture for our creeping system, first we
fitted the event rate functions by the Omori law determining the value of the
parameters $A$, $c$, and $p$. Then the analytic form equation (\ref{eq:omori})
of $n\left(t\right)$ with the numerical parameters was plugged into equation
(\ref{eq:wait_nhpp}) and the integral was calculated numerically taking into
account the load dependent lifetime $t_f\left(\sigma_0\right)$ of the sample
\cite{21,22}, as well. In Fig.\ \ref{fig:waiting_time} an excellent agreement
can be observed between the waiting time distributions obtained from the
simulations $P\left(T\right)$ and the analytic prediction
$P_{NHPP}\left(T\right)$ of equation (\ref{eq:wait_nhpp}). An important
consequence of the above results is that the lower cutoff $T_l$ of
$P\left(T\right)$ can be obtained from the saturation event rate $T_l=1/A$,
which does not depend on the external load. The upper cutoff $T_u$ is determined
by the other time scales as $T_u=\frac{1}{A}\left(\frac{t_f}{c}\right)^p$.
Inserting the Basquin law of creep life $t_f \propto \sigma_0^{-\gamma}$
reproduced by our model \cite{21,22} and the linear load dependence of  it
follows that the upper cutoff $T_u$ scales with the external load as $T_u \sim
\sigma_0^{-\left(1+\gamma\right)}$. The results show that details of the damage
process control the global time scale of rupture, however, they do not affect
the Omori and waiting time exponents. Power law distributions of interoccurrence
times in fracture phenomena are usually considered to be the fingerprint of
correlations between consecutive events \cite{1,2,3,4,5,6,7,8,9,10,11,12,13,14}.
Our analysis revealed that for creep phenomena this is not necessarily true, the
global acceleration of a heterogeneous system can lead to power law distributed
inter-event times without any local correlations.

The comparison of theoretical results to the experimental findings and different
types of measurements to each other can be problematic because in laboratory
experiments the time series of bursts is never complete: small size bursts
generate only low amplitude signals which may fall in the range of background
noise \cite{1,2,3,4,5,6,7,8,9,10,11}. Devices also have a finite time resolution
resulting in a deadtime of detection, during which bursts generated at different
positions in space cannot be distinguished from each other
\cite{12,3,4,5,6,7,8,9,10,11,15,30}. Incompleteness of the time series,
especially in field observations, may also be caused by the fact that recording
does not start exactly at the time when the load was set. Hence, the beginning
of the time series is missing and the measurement is more focused on the
vicinity of the failure point where intensive crackling occurs
\cite{1,2,3,4,32,33,34}. 

\begin{figure}
\begin{center}
\epsfig{bbllx=0,bblly=480,bburx=1055,bbury=780, file=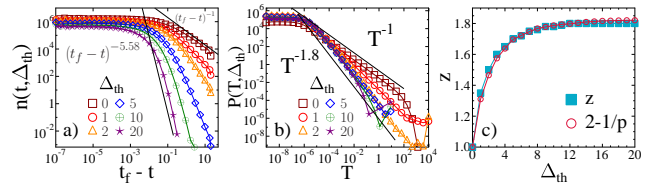,
width=8.2cm}
\caption{Event rates $(a)$ and waiting time distributions $(b)$ at different
detection thresholds $\Delta_{th}$. $(c)$ Comparison of the exponent
$z\left(\Delta_{th}\right)$ of the waiting time distributions obtained
numerically with the NHPP prediction equation (\ref{eq:z_p}).
}
\label{fig:omori_hatter}
\end{center}
\end{figure}

In order to capture the effect of the detection threshold of the measuring
equipment in the data evaluation, we introduced a threshold value $\Delta_{th}$
for the size of bursts $\Delta$ , i.e. bursts with size $\Delta \leq
\Delta_{th}$  are ignored in the time series. Since the size of bursts increases
when approaching global rupture (see Fig.\ \ref{fig:timeseries}), the detection
threshold removes events typically at the beginning of the time series
decreasing the rate of events in this regime. It can be observed in Fig.\
\ref{fig:omori_hatter}$(a)$ that as $\Delta_{th}$ increases the functional form
of the event rate $n\left(t,\Delta_{th}\right)$ remains nearly the same
described by the Omori law equation (\ref{eq:omori}), however, the exponent $p$
monotonically increases with the threshold value $\Delta_{th}$. For the
corresponding waiting time distributions $P\left(T,\Delta_{th}\right)$ in Fig.\
\ref{fig:omori_hatter}$(b)$, the exponent $z$ of the power law regime also
increases with increasing $\Delta_{th}$, however, the NHPP nature of the event
series is preserved at any values of $\Delta_{th}$. For NHPPs the two exponents
$z$ and $p$ have the simple relation \cite{28}
\beq{
z(\Delta_{th}) = 2-1/p(\Delta_{th}),
\label{eq:z_p}
}
which holds with a high accuracy in our system for the numerically determined
exponents at any values of $\Delta_{th}$ (see Fig.\
\ref{fig:omori_hatter}$(c)$). It is important to emphasize that the exponents
increase due to the non-stationary nature of creep rupture: the simultaneous
increase of the rate and size of events towards failure has the consequence that
the finite detection threshold mainly affects the beginning of the time series
resulting in a few long waiting times up to the first bigger bursts. Their
statistics is characterized by a peak or small bump in Fig.\
\ref{fig:omori_hatter}$(b)$, while the rest of waiting times have a steeper
power law distribution. The results demonstrate that the detection threshold has
a dramatic effect on the outcomes of the analysis of crackling time series, just
varying $\Delta_{th}$ practically any values can be obtained for the waiting
time exponent $z$ between $1$ and $2$.

\begin{figure}
\begin{center}
\epsfig{bbllx=0,bblly=450,bburx=710,bbury=750, file=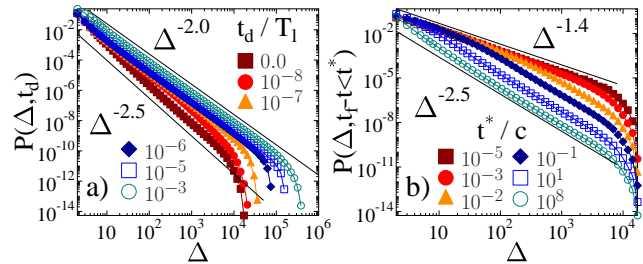,
width=8.2cm}  
\caption{$(a)$ Burst size distributions at $\sigma_0 / \sigma_c = 0.001$ for
different values of the deadtime $t_d$. A crossover is observed from a power law
of exponent $\tau=2.5$ to a lower value $\tau_d=2.0$  $(b)$ Considering events
solely close to failure $P\left(\Delta,t_f-t<t^*\right)$ shows a crossover from
the exponent $2.5$ to $1.4$.
}
\label{fig:holdido_dist}
\end{center}
\end{figure}

The finite time resolution $t_d$ of the detectors has the consequence that
bursts pile up, i.e. since bursts cannot be distinguished within the duration
$t_d$, the size of bursts sums up giving rise to larger event sizes in the time
series. The effect of the deadtime is captured in the data evaluation such that
if an avalanche of size $\Delta_i$ occurred at time $t_i$, all those avalanches
which appeared in the interval $t_i<t<t_i+t_d$ are added to $\Delta_i$. In Fig.\
\ref{fig:holdido_dist}$(a)$ at zero deadtime $t_d=0$ where all avalanches are
distinguished the size distribution $P\left(\Delta,t_d=0\right)$ has a power law
form
\beq{
P(\Delta,t_d=0) \sim \Delta^{-\tau}
\label{eq:deltadist}
}
followed by an exponential cutoff. The value of the exponent $\tau=2.5$ is equal
to the usual mean field burst exponent of FBMs \cite{16,17,18,18,20,21,22}. As
$t_d$ increases the pile up of bursts promotes large events while the small ones
get suppressed. Due to the acceleration of the failure process pile-up gets
dominating in the vicinity of macroscopic failure, hence, in Fig.\
\ref{fig:holdido_dist}$(a)$ the value of $t_d$ is compared to $T_l$ of the
waiting time distribution. As a consequence, the waiting time distributions
hardly change, however, the burst size distribution $P\left(\Delta,t_d\right)$
has a crossover to a power law of a significantly lower exponent $\tau_d=2.0$
showing the higher frequency of large events in the statistics. In Fig.\
\ref{fig:holdido_dist}$(a)$ pile up becomes dominating already at $t_d/T_l
\approx 0.001$, which shows the importance of the results for real experiments.

Recently, laboratory experiments on earth materials have revealed that the
b-value, i.e. the exponent of the probability distribution of the energy of the
time series of acoustic events of rupture cascades decreases in the vicinity of
failure \cite{31,32,33,34}. To investigate the possibility of an analogous
phenomenon in creep rupture, we constrained the data evaluation to events
occurring in a time interval of duration $t^*$ preceding macroscopic failure and
determined the probability distribution $P\left(\Delta,t_f-t<t^*\right)$. It can
be observed in Fig.\ \ref{fig:holdido_dist}$(b)$ that approaching macroscopic
rupture $t^*<c$, where the largest avalanches are triggered, the burst size
distribution $P\left(\Delta,t_f-t<t^*\right)$ exhibits a crossover: at a
characteristic burst size $\Delta_c$ the exponent of
$P\left(\Delta,t_f-t<t^*\right)$ has a striking change from $\tau=2.5$ to a
surprisingly low value $\tau \approx 1.4$. The value of $\Delta_c$ extends to
the largest avalanche as $t^*/c$ decreases. The crossover is accompanied by the
change of the waiting time distribution, as well: Since in the regime $t^*<c$
the event rate is constant, the power law regime of $P\left(T,t_f-t<t^*\right)$
disappears and the distribution turns to a pure exponential as it is expected
for homogeneous (constant event rate) Poissonian processes \cite{29}.

\section{Discussion}

Acoustic outbreaks generated by nucleating and propagating cracks provide the
main source of information on the microscopic temporal dynamics of creep
rupture. For the understanding of acoustic monitoring data of engineering
constructions and of field measurements on steep slopes or rock walls in
mountains requires the application of statistical physics. Our analysis showed
that time-to-failure power laws of macroscopic quantities such as creep rate
commonly observed in experiments are accompanied by the emergence of Omori type
acceleration of the bursting activity on the microscale. The origin of the Omori
time scale is that the aging system drives itself to a critical state where a
few breakings are sufficient to trigger extended bursts. The Omori law is known
to describe the relaxation of the rate of aftershocks following major
earthquakes \cite{26,27}, and it has also been confirmed for foreshocks when
observed \cite{32}. Our results suggest the interpretation that acoustic bursts
in creep behave like foreshocks of the imminent catastrophe. 

Our investigations unveiled that the evolving time series of crackling events is
the result of an underlying non-homogeneous Poissonian process. It has the
striking consequence that observing power law distributed waiting times in
fracture may not imply the presence of dynamic correlations, up to a large
extent it can be caused by the global acceleration of the system. We showed that
special care should be taken when comparing results of model calculations to
measurements on crackling noise, since the deadtime of devices and the finite
background noise to signal ratio can even affect the measured value of critical
exponents. Varying solely the detection threshold of events, for the
distribution of waiting times any exponents can be obtained between $1$ and $2$
covering the range of experimental results
\cite{1,2,3,4,5,6,7,8,9,10,11,12,13,14,15}. Due to the finite deadtime of
electronics bursts pile up which gives rise to a crossover to a lower exponent
of the size distribution of bursts. Recently, avalanches have been identified
with a high spacial resolution along a propagating crack front using optical
imaging techniques \cite{15,30}. Considering global avalanches in the same
experiment integrates bursts along the front, giving rise to a significantly
lower exponent in agreement with our predictions \cite{15,30}. 

Components of engineering constructions are mainly subject to creep loads
\cite{3}, and creep rupture often lies at the core of natural catastrophes such
as landslides, snow and stone avalanches, as well \cite{31,32,33,34}. We
demonstrated that restricting the measurement to the close vicinity of ultimate
failure, the size distribution of bursts exhibits a crossover to a significantly
lower exponent, which is accompanied by the change of the functional form of the
waiting time distribution. Since the crossover is controlled by the time scale
of the Omori law these results can be exploited for forecasting the imminent
catastrophic failure event. 

Here we focused on the case of long range load redistribution following failure
events. When the load sharing is localized the spatial correlation of failure
events leads to the emergence of a propagating crack front. The load accumulated
along the crack front gives rise to an overall acceleration of the failure
process again with a non-homogeneous Poissonian character. However, at short
time scales correlated clusters of events may arise inside the time series. The
results imply the interesting question to clarify when studying burst time
series under creep whether there is anything in the dynamics beyond
non-homogeneous Poissonian processes.

\section{Methods}

We use a generic fiber bundle model \cite{16,17,18,19,20} of the creep rupture
of heterogeneous materials which has succesfully reproduced measured creep
behavior \cite{21,22}. The sample is discretized in terms of a bundle of $N$
parallel fibers having a brittle response with identical Young modulus $E$. The
bundle is subject to a constant external load $\sigma_0$ below the fracture
strength of $\sigma_c$ the system. It is a crucial element of the model that the
fibers break due to two physical mechanisms: immediate breaking occurs when the
local load $\sigma$ on fibers exceeds their fracture strength $\sigma_{th}^{i}$,
$i=1,\ldots , N$. Time dependence is introduced such that those fibers, which
remained intact, undergo an aging process accumulating damage $c\left(t\right)$
\cite{21,22}. The rate of damage accumulation $\Delta c\left(t\right)$  is
assumed to have a power law dependence on the local load $\sigma\left(t\right)$
of fibers $\Delta c\left(t\right) = a\sigma\left(t\right)^{\gamma}\Delta t$,
where $a$ is a constant and the exponent $\gamma$ controls the time scale of the
accumulation process. Fibers can tolerate only a finite amount of damage so that
when the total damage $c\left(t\right)$ accumulated up to time $t$ exceeds a
local damage threshold $c_{th}^i$ the fiber breaks. The two breaking thresholds
$\sigma_{th}^i$ and $c_{th}^i$, $i=1,\ldots , N$ of fibers are independent
random variables which are for simplicity uniformly distributed between $0$ and
$1$. After each breaking event the load dropped by the broken fiber is equally
redistributed over the remaining intact ones \cite{21,22}.

The separation of time scales of slow damage and of immediate breaking leads to
the emergence of a bursty evolution of the system: damaging fibers break slowly
one-by- one, gradually increasing the load on the remaining intact fibers. After
a certain number of damage breakings the load increment becomes sufficient to
induce the immediate breaking of a fiber which in turn triggers an entire burst
of breakings. As a consequence, the time evolution of creep rupture occurs as a
series of bursts corresponding to the nucleation and propagation of cracks,
separated by silent periods of slow damaging. The size of burst $\Delta$ is
defined by the number of fibers breaking in a correlated trail, while the
waiting time $T$ between consecutive events is the duration of the damage
sequence of $\Delta_d$ breakings which triggers the next event. Macroscopic
failure occurs in the form of a catastrophic avalanche at time $t_f$ which
defines the lifetime of the system \cite{21,22}. Our model reproduces the
Basquin law of creep life, i.e. the lifetime $t_f$ decreases as a power law of
the external load $t_f \propto \sigma_0^{-\gamma}$, where the exponent coincides
with the exponent of damage accumulation $\gamma$ \cite{21,22}.

Computer simulations were carried out in our FBM with $N=10^7$ fibers averaging
over $1000$ samples for each parameter set except for Fig.\ \ref{fig:timeseries}
where intensionally a small bundle of $N=10^5$ fibers were considered. The
exponent $\gamma$ of the damage law mainly sets the global time scale of the
evolution, hence, the simulation results are presented only for $\gamma=1$.
Other parameters of the model are fixed as $E=1$ and $a=1$ \cite{21,22}.

\section{acknowledgments}
This work was supported by the projects TAMOP-4.2.2.A-11/1/KONV-2012-0036,
TAMOP-4.2.2/B-10/1-2010-0024, TÁMOP4.2.4.A/2-11-1-2012-0001, OTKA K84157, and
ERANET\_HU\_09-1-2011-0002.

\end{document}